\documentclass[twocolumn,prl]{revtex4}
\usepackage{amssymb}
\usepackage{graphicx}
\usepackage{enumerate}
\begin{document}

\title{Direct Use of Secret Key in Quantum Cryptography}
\author{Horace P. Yuen$^*$\\
Department of Electrical Engineering and Computer Science\\
Department of Physics and Astronomy\\
Northwestern University, Evanston, IL 60208}

\begin{abstract}
For single-photon quantum key generation, it is
shown that the use of a shared secret key
extended via a pseudo-random number generator may simultaneously
enhance the security and efficiency of the cryptosystem. This
effect arises from the intrinsic performance difference between
quantum detectors with versus without knowledge of the key, a
purely quantum effect and a new principle for key generation. No
intrusion level estimation is needed and the method is directly
applicable to realistic systems with imperfections. Quantum direct encryption is also made possible by such use of a secret key. 
\end{abstract}

\maketitle
Considerable progress has been made in the field of quantum
cryptography \cite{gisin02}, mainly in quantum key generation
(QKG) where fresh keys can in principle be generated between two
users, Adam and Babe, with information-theoretic security against
an active attacker Eve for single-photon systems in the asymptotic
limit. This involves an advantage creation process during which
the users establish, via quantum effects, a better communication
channel between themselves for their selected data and measurement
results compared to what Eve may get. They then employ error
correction and privacy amplification to obtain a final generated
key that Eve has little information about, which can be utilized
for other purpose such as the secret key in a direct encryption
system. However, the problems of quantifying the actual security
level and efficiency of a realistic finite bit-sequence QKG system
remain unsolved. One source of the difficulty is the statistical
fluctuation in a concrete finite bit-sequence cryptosystem. Thus, during the process of
intrusion-level estimation (ILE) in which the users create their
advantage over Eve, false alarms may occur that adversely affect
the final security and efficiency tradeoff of the overall
cryptographic protocol \cite{yuen03}. Also, the side information
leaked to Eve during error correction and privacy amplification is
hard to estimate accurately \cite{note1}, and accounting for them
via simple bounds \cite{cachin97} often lead to precipitous drops
in efficiency. There are also well-known problems from system
imperfections, and other problems to be explained elsewhere.

In this paper, the direct use of a shared secret seed key expanded
via a pseudo-random number generator (PRNG) to a running key that
determines the choice of basis in standard 4-state BB84 is
proposed and analyzed against individual and collective attacks.
In general, such use of a shared secret key allows the users to
create advantage via the optimal quantum receiver principle:- Even
if a full copy of the quantum state is available to Eve, the
optimal quantum detector that the users may employ with knowledge
of the secret key performs better than the optimal detector Eve
may employ by a quantum measurement \emph{without} knowledge of
the key even if she may subsequently use the key on her
measurement result. This performance difference does \emph{not}
exist if the system is described classically. This
\emph{principle}, called KCQ (keyed communication in quantum
noise) key generation, constitutes a new and more powerful
advantage creation mechanism via quantum effect. It was described
in ref \cite{yuen03}, briefly in ref \cite{yuen05}, and mentioned
in ref. \cite{barbosa03} where the use of a PRNG on mesoscopic
coherent states for direct encryption was experimentally
demonstrated that does not involve the above KCQ key generation
principle. With the many possibilities of different
implementations, the power of this principle is illustrated in
this paper for single-photon BB84 and corresponding Ekert
protocols. It will be shown how the security/efficiency of a
concrete BB84 cryptosystem may be improved and quantified. Among
the advantages are the alleviation of the problems mentioned in
the last paragraph. It also makes quantum direct encryption (QDE)
possible, in contrast to standard BB84. Some of the presented results have been
given in ref \cite{yuen03} along with a more detailed 
general
treatment, but the present paper is self-contained.

Consider the standard 4-state single-photon BB84 cryptosystem, in
which each data bit is represented by one of two possible bases of
a qubit, say the vertical and horizontal states $|1\rangle$,
$|3\rangle$ and the diagonal states $|2\rangle$, $|4\rangle$. In
standard BB84, the choice of basis is revealed after Babe makes
her measurements, and the mismatched ones are discarded. It has
been suggested \cite{hwang98} that some advantages obtain when a
secret key is used for basis determination with usual ILE and the
resulting protocol is also secure against joint attacks
\cite{hwang03}. Clearly, no key can be generated after subtracting
the basis determination secret key if a fresh key is used
for each qubit. It was proposed in refs. \cite{hwang98,hwang03}
that a long $m$-bit secret key is to be used in a longer $n$-qubit
sequence with repetition. However, even if such use does not
affect the average information that Eve may obtain, it gives rise
to such an unfavorable distribution that security is seriously
compromised. This is because with a probability $1/2$, Eve can
guess correctly the basis of a whole block of $n/m$ qubits by
selecting the qubits where the same secret bit is used
repetitively. For a numerical illustration, let $n=10^3$ and
$m=10^2$. Then with a probability $2^{-15} \sim 0.3\times10^{-4}$, Eve can
successfully launch an opaque (intercept/resend) attack that gives
full information at the dangerous $15\%$ level \cite{gisin02} on the total bit sequence while yielding
no error to the users. Much stronger attacks can be built upon such partial opaque attacks. In general, the strong correlation from
such repetitive use would seriously affect the appropriate
quantitative security level \cite{note2}, and use of a large $m$ relative to $n$ would severely degrade the efficiency.

This problem is alleviated when a seed key $K_s$ is first passed
through a PRNG to yield a running key $K_r$ that is used for basis
determination, as indicated in Fig.~1. In practice, any standard
cipher running in the stream-cipher mode \cite{menezes} can be
used as a PRNG. Even a LFSR (linear feedback shift register) is
good in the present situation. A LFSR with openly known (minimal)
connection polynomial and initial state $K_s$ generates a
``pseudo-random'' output with period $2^{|K_s|}$ \cite{menezes}.
When a LFSR is used as a (classical) stream cipher, it is insecure
against known-plaintext attack \cite{menezes}, in which Eve would
obtain the seed key from the running key which is itself obtained
from the input data and the output bits. However, there is no such
attack in key generation where Adam picks his data bits randomly.
In an attack where Eve guesses the key before measurement, the system is undermined
completely with a probability of $2^{-|K_s|}$. Since it is
practically easy to have $|K_s|\sim10^3$ or larger in a stream
cipher, such a guessing attack would have a much lower, and indeed
insignificant, probability of success compared to, say, the
guessing attack Eve may launch by guessing the message
authentication key used to create the public channel needed in
BB84.
In contrast to the case without a PRNG, no subset of the data is vulnerable to a guessing attack that would correctly guess a 
subset of the key with a high probability.

\begin{figure} [htbp]
\begin{center}
\rotatebox{-90} {
\includegraphics[scale=0.4]{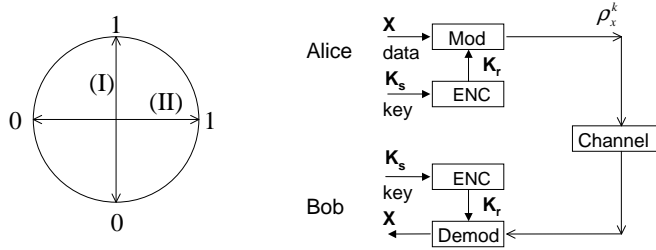}}
\caption{The qb-KCQ scheme. Left -- Two bases, I and II. Right -- Overall
encryption involves modulation with bases determined by a running key $K_r$ generated
from a seed key $K_s$ via an encryption mechanism denoted by the box ENC.}
\end{center}
\end{figure}


Next, we show that the seed key has complete information-theoretic
security against ciphertext-only attacks. Let $\rho_x^k$ be the
quantum state corresponding to data sequence $x=x_1 \cdots x_n$
and running key $k_1 \cdots k_n$. For attacking the seed key $K_s$
or running key $K_r$, the quantum ciphertext reduces to $\rho^k =
\sum_x {p_x \rho^k_x}$ where $p_x$ is the apriori probability of
$x$. In our KCQ approach, we grant a full copy of the quantum
state to Eve for the purpose of bounding her information
\cite{note3}. By an optimal measurement on the qubits, Eve's
probability of correctly identifying $k$ (i.e., $K_r)$ and then
$K_s$ may be obtained via $\rho^k$. Since each qubit is modulated
by its own corresponding data bit, we have $\rho_x^k =
\rho^{k_1}_{x_1} \otimes \cdots \otimes \rho_{x_n}^{k_n}$. For
uniform data commonly assumed for key generation, the $X_i$ are
independent, identically distributed (i.i.d.) Bernoulli random
variables with equal probabilities. Thus each $\rho^{k_i}=
{I_i}/2$ after averaging over $x_i$ for any value of $k_i$, with
resulting $\rho^k = \bigotimes_{i=1}^n {I_i}/2$ completely
independent of $k$. So Eve can obtain no information on $K_r$ or
$K_s$ at all even if she possesses a full copy of the quantum
signal.

Next, we quantify the minimum security level against collective
attacks on the random data, for which Eve is assumed to have a
full copy of the quantum signal. By ``\emph{collective attack}''
we mean the situation where Eve performs a constant qubit-by-qubit
measurement on her (fictitious) full copy in the absence of any
knowledge on $K_s$ or $K_r$, but may employ collective classical
processing of the measurement results to take into account
correlations induced by $K_s$. This is analogous but different
from the usual ``collective attacks'' in BB84, because there is no
question of probe setting in the present case. Since the term
``individual attack'' in BB84 does not include collective
classical processing, our use of the term ``collective attack'' is
appropriate and allows the further generalization to joint
measurements in the most general case of ``joint attacks''
\cite{note4}. Whatever the terminology,  $K_s$ or $K_r$ is actually never revealed to Eve so
that all her knowledge of the secret key must come from her
quantum measurements. Practically, so long as Eve does not have
long-term quantum memory, she would need to measure the qubits
even if she could obtain $K_r$ at a future time, which is actually
impossible if the key is never used again.

Nevertheless, solely for the purpose of lower bounding Eve's
information (which is difficult to estimate otherwise because of
the correlations introduced by $K_s$ among the qubits), here we
conceptually grant Eve the actual $K_s$, and hence $K_r$, after
she made her measurements. Our KCQ principle of key generation via
optimal quantum receiver performance with versus without knowledge
of $K_s$ is easily seen to work here: Even with a full copy of the
quantum state Eve is bound to make errors in contrast to the
users. Indeed, her optimal measurement can be found by
parametrizing an arbitrary orthogonal basis which she measures,  and
optimizing the parameters assuming that $K_s$ is then granted to
her. It is readily shown that general POVM measurements reduce
to othogonal ones in this optimum binary quantum decision problem on
a qubit. Not surprisingly, her optimal error rate is $\sim
0.15$ and is obtained via the ``Breidbart'' basis \cite{bennett92}
well-known in BB84, for which one basis vector bisects the angle
between $|1\rangle$ and $|2\rangle$ or $|2\rangle$ and $|3\rangle$
depending on the bit assignment, and also in this case by the
basis obtained by rotating the Breidbart basis by $\pi/4$. 

It may be observed that security against joint attacks seems to be
also obtained, for the same reason that Eve would make many more
errors except with an exponentially  small (in $|K_s|$)
probability, while the users make none in the noiseless limit.
However, there are problems of counting the secret key bits used
and the appropriateness of the entropy condition \cite{note2}. A
detailed treatment of joint-attack security will be given
elsewhere.

A key verification phase is to be added in a complete protocol
after error correction and privacy amplification, which should
always be employed in key generation for practical reasons. For
this purpose, one may produce a short hashed version $K'_g$ of the
final generated key $K_g$, and encrypt $K'_g$ with, say, one-time
pad from either A to B or B to A. The hash function used may also
be keyed, as usually assumed in BB84 for the message
authentication that creates the ``unjammable'' public channel
guaranteeing unconditional security [10,Ch 9] even if no other
authentication is used. The total number of secret key bits
expended during protocol execution, including $K_s$, is to be
subtracted from the generated key. For key generation protocols
with nonzero key generation rate, including our qubit KCQ scheme
above under collective attacks, such cost may be made arbitrarily
small in rate for large $n$. Notice that in the present case it
does \emph{not} matter what Eve did in her interference during the
protocol execution. As long as $K_g$ is verified, it is correct
except with an exponentially small (in the length of the
verification key length $|K_v|$) probability, and Eve only knows
$K_g$ with a similar probability. Since her information is bounded
with a full copy of the quantum state already granted to here,
there is no need for intrusion-level estimation to ascertain her
information as a function of her disturbance.

The above scheme may be generalized in many obvious ways. One is
to allow $M$ possible bases on the qubit Bloch sphere. This would
increase security without compromising efficiency as in the BB84
case, because there is no mismatched qubits to throw away and
there is no need to communicate openly what bases were measured.
It is readily shown that in the limit $M \rightarrow \infty$,
Eve's error rate goes to the maximum value 1/2 for collective
attacks also, not just individual attacks \cite{yuen03}. Also, the
scheme evidently works in the \emph{same} way for Ekert type protocols
that involve shared entangled pairs. Furthermore, the same
principle may be employed for coherent-state systems with
considerable number of photons \cite{yuen03,yuen05}, although the
corresponding security analysis is more involved and would be
treated in detail elsewhere.

In the present approach, error correction may be carried out by a
forward error correcting code and the resulting performance
analysis is not burdened by the need to consider Eve's probe and
whether she may hold it with quantum memory. If the channel is
estimated to have an error rate $p_c$ below 15\%, advantage is created
against collective attacks as shown above, and the existence of a
protocol that yields a net key generation rate may be proved
asymptotically as usual \cite{note5}. This channel error rate
estimation is not for advantage creation, and is not necessary in
two-way interactive error-correcting procedures, because the KCQ
principle already guarantees the users' advantage over Eve. It is
for correcting the users' channel noise and can be carried out at \emph{any}
time in contrast to ILE.
Such a channel characterization is always needed in any communication line. 

In particular, as long as  $p_c$ is below
the threshold $\sim 0.15$, the users could employ an error
correcting  code with rate $R$ such that 
\begin{equation}
1-h_2(p_c) > R > 1-h_2(0.15), 
\end{equation}
where $h_2(\cdot)$ is the binary entropy function and $1-h_2(\cdot)$ is the capacity of the corresponding (BSC) channel. The second 
inequality in (1) ensures that Eve could not get at the data because the code rate $R$ exceeds her capacity. Under the first inequality 
in (1) or a tighter one for concrete codes, the users can correct the channel errors.

For concrete protocols there is
the general problem of assuring that the side information Eve has on
the error correction and privacy amplification procedures would
not allow her to obtain too much information on $K_g$. Under the
(unrealistic) assumption that only individual classical processing
of each qubit measurement result is made, which is the i.i.d.
assumption underlying many BB84 security analyses, Eve's Renyi
entropy, Shannon entropy, and error rate are simply related.
Quantitative results can then be easily stated as usual. For
collective and general attacks there is the problem of estimating
the Renyi entropy for applying the privacy amplification theorem
\cite{note1}. With ILE in concrete BB84 protocols, this
Renyi 
entropy estimate has never been carried out, with the difficulty
compounded by finite-$n$ statistical fluctuations. The problem is
much alleviated for KCQ protocols for which a single
quantum copy is already granted to Eve for quantitatively bounding
her information. In particular, all the side information from error correction is accounted for by the second inequality of (1). The 
correct way to deal with the privacy amplification problem and a general treatment using a
proper security criterion would be detailed elsewhere.

The complete key generation protocol, to be called qb-KCQ, is
given schematically as follows:

\begin{enumerate}[(i)]
\item Adam sends a sequence of $n$ random bits by a sequence of
$n$ qubit product states, each chosen randomly among two
orthogonal bases via a running key $K_r$ generated by using a PRNG
on a seed key $K_s$ shared by Babe.
\item An error control and privacy amplification procedure is employed by
the users to correct their channel errors and obtain a final
generated key $K_g$, while assuring , e.g.~under (1), that even after all the
associated side information and a full copy of the quantum state
is granted to Eve, errors remain for her so she has little
information on $K_g$.
\item The users employ key verification as in message
authentication to verify that they share the same $K_g$.
\end{enumerate}

The above protocol can be easily modified for performing direct
data encryption. Instead of randomly chosen bits, Adam sends the
data out as in (i) with error control coding but no privacy
amplification. The key verification (iii) becomes just the usual
message authentication. The efficiency and security of such qb-QDE
systems and their extensions for protecting against various attacks will be detailed in the future.
Note that QDE is \emph{not} possible with BB84 or its secret-key
modification in refs \cite{hwang98,hwang03}, the former because of
key sifting, the latter because of the serious security breach of
Eve getting correctly a whole block of $n/m$ data bits with
probability $1/2$ described above.

In sum, the advantages of qb-KCQ over the corresponding
single-photon BB84 key generation are:

\begin{enumerate}[(1)]
\item Efficiency is increased in that there are no wasted qubits and no
need for public communication except for key verification, while
security is increased, especially for large
number of possible bases.

\item No intrusion level estimation is required, thus no
false-alarm problem or any statistical fluctuation problem
associated with such estimation.

\item The security/efficiency analysis is unaffected even for a
multi-photon source whose output state is diagonal in the
photon-number representation \cite{note6}, as a full copy of the
single-photon state is already granted to Eve for bounding her
information.

\item The security/efficiency quantification is similarly extended
to realistic lossy situations, while new analysis not yet
performed is otherwise needed to take into account, e.g., attacks
based on approximate probabilistic cloning \cite{fiurasek04}.

\item The security/efficiency analysis is also similarly extended to
include any side information for a finite-$n$ protocol, with no
question of holding onto the probes.

\item There are practical advantages in reducing the number of random data bits needed by Adam and photon counters needed by Babe in an 
experimental implementation.

\item Direct encryption without going through key generation
first may be employed, which is impossible for BB84.
\end{enumerate}

In conclusion, a new approach involving shared secret key and the
KCQ principle for obtaining secure and efficient qubit key
generation has been described that dispenses with intrusion level
estimation and allows quantum direct encryption. Since a
shared secret key is already needed for message authentication
during protocol execution in current quantum key generation
protocols, one should also employ a seed key expanded to a running
key as in our qb-KCQ scheme to enhance simultaneously the
security and efficiency of the cryptosystems.

{\bf Acknowledgements --}
I would like to thank R.~Nair, W.-Y.~Hwang, E.~Corndorf, P.~Kumar, C.~Liang, and K.~Yamazaki for useful
discussions. This work was supported by DARPA grant
F30602-01-2-0528.
 
$^*$E-mail: yuen@eecs.northwestern.edu.


\begin{thebibliography}{10}
\bibitem{note2}
Even if the average quantity is a small percentage of the total,
the underlying probability distribution may be so skewed that it
is unacceptable for security guarantee. For an example involving
entropy, see Section III.C in ref. [4].
\bibitem{note3}
She is given only a single copy, not a ``cloning machine''. In
single-photon BB84, one such copy is enough to completely
undermine its security.
\bibitem{note4}
Note that just non-identical qubit-by qubit probes is already classified as
a joint attack in the BB84 context.
\bibitem{note5}
Note that no concrete protocol with a specific code is exhibited in such asymptotic
proofs.
\bibitem{note6}
We assume that the multi-photon nature of the source does not
permit an unambiguous state determination (USD) type attack that
is more powerful than the availability of a full copy of the
quantum state.
\end{thebibliography}
\end{document}